# Oppositely Directional Coupler: Example of the Forward Backward Waves Interaction in the Metamaterials

A.I. Maimistov[1,2], E.V. Kazantseva[2,3]

[1]Department of Solid State Physics and Nanosystems, National Research Nuclear University Moscow Engineering Physics Institute, 115409 Moscow, Russia

[2]Department of Physics and Technology of Nanostructures, Moscow Institute of Physics and Technology 141700, Dolgoprudnyi, Moscow region, Russia

[3] Joint Institute for High Temperatures, Russian Academy of Sciences, 125412 Moscow, Russia

**Abstract**. We consider the coupled electromagnetic waves propagating in a nonlinear coupler and in nonlinear waveguide array, which consists of alternating waveguides of positive and negative refraction indexes. The forward wave and backward wave interaction is realized in these devices. Gap solitons in a nonlinear oppositely directional coupler with one channel or both channels fabricated from nonlinear medium having negative refraction index are discussed. Generalization of the usually waveguide array is zigzag array. Due to zigzag configuration there are interactions between both nearest and next nearest neighboring waveguides exist. The system of evolution equations for coupled waves has the steady state solution describing the electromagnetic pulse running in the array. Numerical simulation demonstrates robustness of these solitary waves.

## 1 Introduction

If the phase velocity and the Pointing vector of the incident electromagnetic wave are directed in the same direction but the phase velocity and the Pointing vector of the refracted electromagnetic wave are opposite directed, than refraction angle is negative one. The Snell's formula can be used in this case, if the refractive index is considered as negative index. This phenomenon is referred to as negative refraction. Antiparallel orientation of the phase velocity ($v_{ph}$) and the Pointing vector ($S$) was first discussed in [1, 2]. In [3] it was indicated that antiparallel orientation of $v_{ph}$ and $S$ results in negative refraction. Subsequently, this idea was developed by Mandelstam in [4]. It has been predicted that when the real parts of the dielectric permittivity and magnetic permeability in the medium simultane-ously take on negative values in some frequency range, antiparallel orientation of $v_{ph}$ and $S$ occurs [5,6] and the property of negative refraction appears [7].

The existence of the media characterized by negative refractive index (NRI) was demonstrated experimentally first in the microwave and then in the near-infrared ranges [8-12]. Reviews of the properties NRI materials are presented in [13-16].

The present technological level does not yet allow for the fabrication of 3D materials of sufficient size and small enough losses for experiential verification of the effects described above. However, a considerable effort aimed at loss reduction and improvement of nanofabrication technology gives hope that the considered device will be manufactured.

It is well known that two closely located waveguides can be coupled due to the tunneling of light from one waveguide to the other. A coupler using tunneling, fabricated from materials with a positive refractive index (PRI), preserves the direction of light propagation, and for this reason it is named a *directional coupler*. It is much used device in integral optics.

If one of the waveguides of the coupler is fabricated from a material with a negative refractive index, this device is taking new features. The radiation entering one waveguide leaves the device through the other waveguide at the same end but in the opposite direction. For this reason, this device can be called the *oppositely directional coupler*. The properties of this coupler and different generalization of oppositely directional couplers will be discussed here.

## 2. Oppositely directional coupler

The principal property of the oppositely directional coupler is governed by the spectral features of linear wave. Unlike wave in usual directional coupler the spectrum of waves in the opposite directional coupler has the forbidden zone (i.e., gap). In this case the coupler acts like to distributed mirror.

### 2.1. Linear oppositely directional coupler

The electric field of an optical wave propagating in linear directed coupler in the positive $z$ direction can be represented as follows

$$E(x,y,z;t) = \sum_{J=1,2}\sum_{m} A_m^{(J)}(z,t)\Psi_m^{(J)}(x,y)\exp[-i\omega_0 t + i\beta_m^{(J)}z], \quad (1)$$

The mode function for a particular $m$-th mode of channel $J$ is denoted by $\Psi_m^{(J)}(x,y)$, and $A_m^{(J)}$ is a slowly varying envelope of the electric field corresponding to this mode. Parameters $\beta_m^{(J)}$ are propagation constants. Omitting the details we can write the general coupled equations which are governed by normalized envelopes $Q_J(\zeta,\tau) = A_m^{(J)}(z,t)A_0^{-1}$

$$i\widetilde{k}_1 \frac{\partial Q_1}{\partial z} + i\frac{1}{v_{g1}}\frac{\partial Q_1}{\partial t} + K_{12}Q_2 \exp\{+i\Delta\beta z\} = 0,$$

$$i\tilde{k}_2 \frac{\partial Q_2}{\partial z} + i\frac{1}{v_{g2}} \frac{\partial Q_2}{\partial t} + K_{21} Q_1 \exp\{-i\Delta\beta z\} = 0 \ .$$

The coefficients $K_{12}$ and $K_{21}$ are the coupling constants between neighboring waveguides. The phase mismatch is taking into account by $\Delta\beta = \beta_m^{(2)} - \beta_m^{(1)}$. $v_{g1}$ and $v_{g2}$ are group velocities. If in these equations we put $\tilde{k}_1 = +1$ and $\tilde{k}_2 = -1$ in corresponding to PRI and NRI cases respectively, we get the mathematical models describing the linear oppositely directional coupler.

It is suitable to introduce new variables

$$q_1 = \sqrt{K_{21}} Q_1 \exp\{-i\Delta\beta z/2\}, \quad q_2 = \sqrt{K_{12}} Q_2 \exp\{+i\Delta\beta z/2\} \ . \tag{2}$$

and the new variables according to following formulae $\zeta = z/L_c$, $\tau = t_0^{-1}(t - z/V_0)$, where $\delta = \Delta\beta L/2$ and

$$L_c = (K_{12} K_{21})^{-1/2}, \quad t_0 = L(v_{g1} + v_{g2})/2v_{g1}v_{g2}, \quad V_0 = (v_{g2} - v_{g1})/2v_{g1}v_{g2} \ .$$

The system of the linear equations is rewritten as

$$i\left(\frac{\partial}{\partial\zeta} + \frac{\partial}{\partial\tau}\right) q_1 - \delta q_1 + q_2 = 0, \quad i\left(\frac{\partial}{\partial\zeta} - \frac{\partial}{\partial\tau}\right) q_2 + \delta q_2 - q_1 = 0 \ . \tag{3}$$

Using the Fourier transformation we can find the dispersion relation for the harmonic waves $\omega = \delta \pm \sqrt{1 + k^2}$. Thus, this spectrum of harmonic waves has the gap $\Delta\omega = 2$. It should be pointed out that in the case of the PRI medium of both channels the gap is absent.

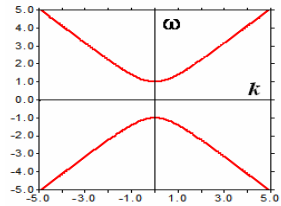

**Fig.1** The dispersion relation for linear waves in oppositely-directional couplers.

## 2.2. Nonlinear oppositely directional coupler

If the channels of directed coupler are prepared from a nonlinear medium, the system of the general coupled equations describing the nonlinear opposite directional coupler (NLODC) will contain additional terms. If normalized variables are used then the system of equations reads [17-19]

$$i\left(\frac{\partial}{\partial \zeta}+\frac{\partial}{\partial \tau}\right)q_1 - \delta q_1 + q_2 + r_1 |q_1|^2 q_1 = 0, \quad (4)$$

$$i\left(\frac{\partial}{\partial \zeta}-\frac{\partial}{\partial \tau}\right)q_2 + \delta q_2 - q_1 - r_2 |q_2|^2 q_2 = 0. \quad (5)$$

There parameters $r_1$ and $r_2$ are the nonlinearity measure for each waveguide [18]. By using real variables $q_{1,2} = a_{1,2}\exp(i\varphi_{1,2})$, one can obtains from (4)-(5) the system of the real variables equation.

$$\left(\frac{\partial}{\partial \zeta}+\frac{\partial}{\partial \tau}\right)a_1 = a_2 \sin\Phi, \quad \left(\frac{\partial}{\partial \zeta}+\frac{\partial}{\partial \tau}\right)a_1 = a_2 \sin\Phi,$$
$$\left(\frac{\partial}{\partial \zeta}+\frac{\partial}{\partial \tau}\right)\varphi_1 = -\delta + \frac{a_2}{a_1}\cos\Phi + r_1 a_1^2, \quad (6)$$
$$\left(\frac{\partial}{\partial \zeta}-\frac{\partial}{\partial \tau}\right)\varphi_2 = \delta - \frac{a_1}{a_2}\cos\Phi - r_2 a_2^2,$$

where $\Phi = \varphi_1 - \varphi_2$.

**CW-limit of nonlinear opposite directional coupler**

Let be $\delta = 0$. Here the NLODC of normalized length $l$ will be considered [17]. From equations (6) the two integrals of motion result

$$a_1^2 - a_2^2 = c_0^2, \quad (7)$$

$$4\sqrt{a_1^2 - c_0^2}\cos\Phi = (r_1 + r_2)a_1^3 - 2r_2 c_0^2 a_1, \quad (8)$$

where constant $c_0$ is defined by boundary condition $a_2(l)=0$. Using these expressions one can reduce the (6) to equation for $a_1$ that can be solved. Finally, amplitudes of the directed waves are

$$a_1^2(\zeta) = c_0^2 \frac{\mathrm{dn}[2(\zeta-l)/m, m]+1}{2\,\mathrm{dn}[2(\zeta-l)/m, m]}, \quad a_2^2(\zeta) = c_0^2 \frac{\mathrm{dn}[2(\zeta-l)/m, m]-1}{2\,\mathrm{dn}[2(\zeta-l)/m, m]}. \quad (9)$$

The parameter $c_0$ is defined now by transcendental equation

$$a_0^2 = c_0^2 \frac{\mathrm{dn}[2l/m, m]+1}{2\,\mathrm{dn}[2l/m, m]} \quad (10)$$

Reflection coefficient of the NLODC is

$$\Re = 1 - a_1^2(l)/a_0^2 = 1 - \frac{c_0^2}{a_0^2} = 1 - \frac{2\,\mathrm{dn}[2l/m, m]}{1+\mathrm{dn}[2l/m, m]} = \frac{1-\mathrm{dn}[2l/m, m]}{1+\mathrm{dn}[2l/m, m]}. \quad (11)$$

The elliptic function dn(z,m) is periodically variable from unit to some positive value that is less unit. Hence, there are value of the $a_0^2$, such that reflection coefficient is zero. From (9) and (10) one can find the dependence output power $a_1^2(l)$ vs input power $a_0^2$. Example of this dependence is represented by plot in Fig.2. There are some interval of $a_0^2$, where one value of the input power corresponds two value of the output power. (In really, there are two stable values and one unstable value.) This phenomenon is referred as bistability. The bistability is famous phenomenon in nonlinear optics [18].

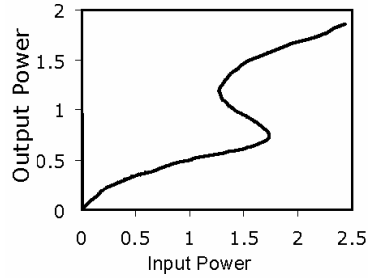

**Fig.2** Bistable behavior of the NLODC [17].

The nonlinear transmission properties of nonlinear oppositely directional coupler with one waveguide made of positive index material and the other waveguide made of negative index material while only one of the waveguide is nonlinear has been considered in [19,20]. In these papers the phase mismatch effect was taken into account. The effect of nonlinearity and mismatch on the multistable behavior for this coupler was studied.

**Gap solitons in opposite directional coupler**

In [21] the solitary wave propagation was considered in frame of the system of equations (4)-(5). In the case of $r_2 = 0$ the solitary wave propagation was investigated in [22]. The existence of the steady state solitary waves was found in both cases. The robustness of these solitary waves in relation to small perturbation has been demonstrated. The system of equations (4)-(5) does not belong to the class of completely integrable equations. Hence the solution of these equations does not represent true soliton. However, we denote them as gap solitons by analogy with gap solitons in nonlinear periodic structures.

To consider the solitary steady state waves in NLODC we have start from the equations (4)-(5) or (6). Suppose that solutions of these equation depend only on single variable $\xi = (\zeta + \beta\tau)\sqrt{1-\beta^2}$, with free parameter $\beta$. Let be $\sqrt{1+\beta}a_1 = u_1$ and $\sqrt{1-\beta}a_2 = u_2$. System of equations (6) after some transformations takes the following form

$$\frac{\partial}{\partial \xi}u_1 = u_2 \sin\Phi, \quad \frac{\partial}{\partial \xi}u_2 = u_1 \sin\Phi, \qquad (12)$$

$$\frac{\partial}{\partial \xi}\Phi = -\frac{2\delta}{\sqrt{1-\beta^2}} + \left(\frac{u_1}{u_2} + \frac{u_2}{u_1}\right)\cos\Phi + \theta_1 u_1^2 + \theta_2 u_2^2, \qquad (13)$$

where

$$\theta_1 = r_1\left[(1+\beta)\sqrt{(1-\beta)/(1+\beta)}\right]^{-1}, \quad \theta_2 = r_2\left[(1-\beta)\sqrt{(1+\beta)/(1-\beta)}\right]^{-1}.$$

Let us consider the case of zero mismatch $\delta = 0$. Solitary wave corresponds with the following boundary condition $a_{1,2} \to 0$ at $\xi \to \pm\infty$. This system of equations has the following integrals of motions correlated with boundary condition under consideration

$$u_1^2 - u_2^2 = 0, \quad 4\cos\Phi + \varepsilon\theta u_1^2 = 0,$$

where $\theta = \theta_1 + \theta_2$, $\varepsilon = \pm 1$. By the use these relations the analytical solution of the system of equations (12)-(13) can be obtained. This solution describes the coupled pair of the forward and backward solitary wave propagating as single wave packet over both waveguides, i.e., gap soliton. Real amplitudes and phases for gap soliton localized at point $\xi_0$ are represented by following expressions [21]

$$a_1^2(\xi) = \frac{4}{\theta(1+\beta)\cosh 2(\xi-\xi_0)}, \qquad a_2^2(\xi) = \frac{4}{\theta(1-\beta)\cosh 2(\xi-\xi_0)}, \qquad (14)$$

$$\varphi_1(\xi) = (1 - 4\theta_1/\theta)\arctan e^{2(\xi-\xi_0)}, \qquad \varphi_2(\xi) = (1 - 4\theta_2/\theta)\arctan e^{2(\xi-\xi_0)} - \pi/2.$$

The negative value of the parameter $\beta$ corresponds to solitary wave which propagates in the direction of the axis $\xi$. The solitary wave characterized by positive value of the parameter $\beta$ propagates in the opposite direction. Large amplitudes of the solitary waves correspond to large positive values of the parameter $\beta$ (which determines pulse velocity). When parameter $\beta$ is negative the gap solitons with smaller values of the parameter $\beta$ correspond to smaller amplitudes, however the absolute value of velocity determined by the parameter $\beta$ is larger for less powerful solitary waves.

The gap soliton formation in the NLODC has a threshold character. A small amplitude electromagnetic pulse, introduced into one of the waveguides, is emitted in the opposite direction from the other waveguide. When the amplitude of the input pulse exceeds a certain threshold, then the pair of coupled pulses propagating in both waveguides is formed. Numerical simulation of the gap soliton formation was produced in [23] under condition that $r_2 = 0$, $r_1 = r$.

Gradually increasing incident pulse amplitude $a_1(\zeta = 0)$ it is possible to approach the threshold value of the amplitude $a_{th}$, when steady state pulse propagating along NLODC is formed (Fig.3).

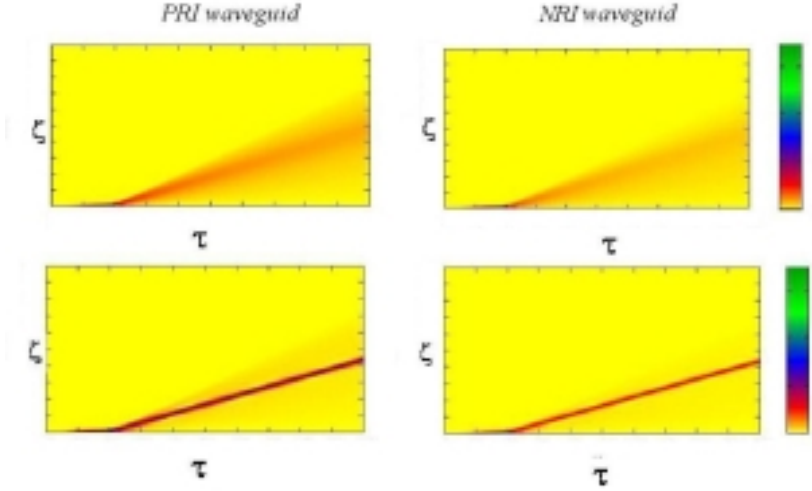

**Fig.3** Gap soliton formation in NLODC. Upper panels are corresponding for incident pulse amplitude that is less threshold value. Lower panels are corresponding for incident pulse amplitude that is over threshold value [23].

Using the expressions (9)-(11) approximately value for $a_{th}$ can be found. It is suggested that soliton is formed on one coupling length, i.e., $l_c = 1$. In this case the coupler will be transparent: $\Re = 0$, that results in $\text{dn}[2/m, m] = 1$. It means that modulus of the elliptical function dn($z,m$) satisfies to equation $m\text{K}(m) = 1$. The transparency of NLODC means that $c_0 = a_{th}$, and one can write for modulus the following formula $m = \left[1+(r\,a_{th}^2/4)^2\right]^{-1/2}$. Under assumption that modulus is a small value the complete elliptic integral $\text{K}(m)$ can be estimated as $\pi/2$. Thus the threshold value of the amplitude $a_{th}$ is defined from the following equation $\left[1+(r\,a_{th}^2/4)^2\right]^{1/2} = \pi/2$. Finely, we have

$$ra_{th}^2 = 4\sqrt{(\pi/2)^2 - 1} \approx 4{,}847 \,. \qquad (15)$$

It is important to emphasize that this expression provides the good estimation for numerical results of [23].

## 3. Alternating nonlinear optical waveguide zigzag array

The optical waveguide array provides a convenient setup for experimental investigation of periodic nonlinear systems in one dimension [24]. Nonlinear optical waveguide arrays (NOWA) are a natural generalization of nonlinear couplers. NOWA with a positive refractive index have many useful applications and are well studied in the literature (see for example [25-27]). If the sign of the index of refraction of one of waveguides in NOWA is positive and the index of refraction of other neighboring waveguide is negative the alternated NOWA will be obtained [16, 28, 29]

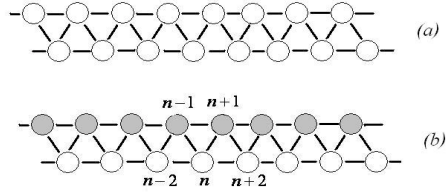

**Fig.4** A schematic illustration of zigzag coupled waveguides array (*a*); and zigzag positive-negative coupled waveguides array (*b*) [32].

Usually the coupling between nearest neighboring waveguides is taken into account. It is correct approximation for strong localized electromagnetic wave in waveguide. However, the coupling between both nearest neighboring waveguides and the next nearest neighboring ones can be introduced by the use of a zigzag arrangement [30, 31] (Fig.4*a*). Let $\vartheta_b$ is an angle between the lines connecting the centers of neighboring waveguides. In a linear array this angle is $\pi$. In a zigzag like array at $\vartheta_b \approx \pi/2$ the coupling between the nearest neighboring waveguides and the next nearest neighboring ones is approximately the same. Nonlinear optical waveguide zigzag arrays can be considered as generation of NOWA.

### 3.1. Model and the base equations

Let us assume that waveguide having numbers $n$ is characterized by positive refractive index (PRI), the nearest neighboring waveguides with numbers $n-1$ and $n+1$ possess negative refractive index (Fig.4*b*). If the electromagnetic radiation is localized in each waveguide the coupled wave theory can be used. In the case of the array is deformed in the form of zigzag, where the angles between the lines connecting waveguides are equal approximately $\vartheta_b \approx 2\pi/3$, interaction between both the nearest neighboring and the next nearest neighboring waveguides will be important. The configuration of these alternating waveguides will be remarked as alternating nonlinear optical waveguide zigzag arrays (ANOWZA).

The system of equations describing the wave propagation in ANOWZA reads as

$$i\left(\frac{\partial}{\partial \zeta}+\frac{\partial}{\partial \tau}\right)q_n + c_1(q_{n-1}+q_{n+1}) + c_2(q_{n+2}+q_{n-2}) + r_1|q_n|^2 q_n = 0, \quad (16)$$

$$i\left(\frac{\partial}{\partial \zeta}-\frac{\partial}{\partial \tau}\right)q_{n+1} - c_1(q_n+q_{n+2}) - c_3(q_{n+3}+q_{n-1}) - r_2|q_{n+1}|^2 q_{n+1} = 0, \quad (17)$$

where $q_n(\zeta,\tau)$ is the normalized envelope of the wave localized in *n*-th waveguide. Coupling between neighboring PRI and NRI waveguides is defined by parameter $c_1$. The $c_2$ ($c_3$) is coupling constant between neighboring PRI (NRI) waveguides. The phase mismatch is taken equal to zero. The all functions $q_n(\zeta,\tau)$, independent variables $\zeta$, $\tau$ and other parameters are expressed in terms of the physical values represented in [32].

### 3.2. Linear waves in alternating waveguide zigzag arrays

I in [32] the asymmetrical ANOWZA was investigated. In this case NRI waveguides are linear ones ($r_2 = 0$). To find the linear wave spectrum we can employ the presentation of the envelopes in the form of harmonic waves.

$$q_n = Ae^{-i\omega\tau+ik\zeta+in\varphi}, \quad q_{n+1} = Be^{-i\omega\tau+ik\zeta+i(n+1)\varphi}.$$

Substitution of this expression in the linear version of the equations (16) and (17) leads to a system of the algebraic linear equations respecting *A* and *B*. This system of linear equations obeys the nonzero solutions if the following condition

$$(\omega+\omega_0)^2 = \gamma_1^2 + (k-k_0)^2$$

will be held. Here the parameters $\gamma_1 = 2c_1\cos\varphi$, $\gamma_2 = 2c_2\cos 2\varphi$, $\gamma_3 = 2c_3\cos\varphi$, and $2\omega_0 = \gamma_2 + \gamma_3$, $2k_0 = \gamma_2 - \gamma_3$ were introduced. Thus, the linear waves in ANOWZA at $r_1 = 0$ and $r_2 = 0$ are characterized by the dispersion relation

$$\omega(k) = \mp\sqrt{\gamma_1^2 + (k-k_0)^2} - \omega_0. \qquad (18)$$

This expression shows that (a) the forbidden zone (gap) in spectrum of the linear waves exist $\Delta\omega = 2|\gamma_1|$, (b) spectrum is shifted along both frequency axis and wave numbers axis, (c) the form of spectrum likes the spectrum for linear oppositely-directional coupler [17,21,22]. The gapless spectrum appears only when condition $\varphi = \pi/2$ is hold. In this case the radiation propagates along waveguides with the same refractive indexes. It should be noted that energy flux between neighboring waveguides is zero.

### 3.2. Nonlinear waves in alternating waveguide zigzag arrays

The intriguing kinds of the nonlinear waves in ANOWZA can be found by the use the following ansatz

$$q_n(\zeta,\tau) = A(\zeta,\tau)e^{in\varphi}, \quad q_{n+1}(\zeta,\tau) = B(\zeta,\tau)e^{i(n+1)\varphi},$$

where $A(\zeta,\tau)$ and $B(\zeta,\tau)$ are the envelopes of the quasi-harmonic waves. It allows to do reduction of the equations (16)-(17) and to obtain the equations

$$i\left(\frac{\partial}{\partial \zeta}+\frac{\partial}{\partial \tau}\right)A+\gamma_1 B+\gamma_2 A+r_1|A|^2 A=0, \quad (19)$$

$$i\left(\frac{\partial}{\partial \zeta}-\frac{\partial}{\partial \tau}\right)B-\gamma_1 A-\gamma_3 B-r_2|B|^2 B=0. \quad (20)$$

These equations have the solutions that describe both steady state wave and spreading waves. The steady state waves correspond to solution of the wave equation depending only one particular variable, for example, $\xi=(\zeta+\beta\tau)\sqrt{1-\beta^2}$. The case of linear NRI waveguides ($r_2=0$) and zero mismatch condition will be discussed now. As above the real variables can be exploited to read the following equations [32]

$$\frac{\partial}{\partial \xi}u_1=u_2 \sin\Phi, \quad \frac{\partial}{\partial \xi}u_2=u_1 \sin\Phi, \quad (21)$$

$$\frac{\partial}{\partial \xi}\Phi=\delta+\left(\frac{u_1}{u_2}+\frac{u_2}{u_1}\right)\cos\Phi+\vartheta u_1^2, \quad (22)$$

where $u_1=\sqrt{1+\beta}|A|$, $u_2=\sqrt{1-\beta}|B|$ are new normalized amplitudes and

$$\delta=\left(\frac{\gamma_3}{\gamma_1}\sqrt{\frac{1+\beta}{1-\beta}}+\frac{\gamma_2}{\gamma_1}\sqrt{\frac{1-\beta}{1+\beta}}\right), \quad \vartheta=\frac{r_1}{\gamma_1(1+\beta)}\sqrt{\frac{1-\beta}{1+\beta}}.$$

From equations (21) and (22) the integrals of motion follow

$$u_1^2-u_2^2=C_1, \quad 4u_1 u_2 \cos\Phi+2\delta u_1^2+\vartheta u_1^4=C_2.$$

The boundary conditions $u_1=u_2=0$ at $\xi\to\pm\infty$ result in following value of theses integrals $C_1=C_2=0$.

Taking into account the integrals of motion the equation (21) and (22) can be solved. The solutions describing the steady state solitary waves are represented by the following expressions

$$|A|^2=\frac{4\Delta^2}{|\vartheta|(1+\beta)\{\cosh[2(\xi-\xi_c)]+\delta/2\}}, \quad (23)$$

$$|B|^2=\frac{4\Delta^2}{|\vartheta|(1-\beta)\{\cosh[2(\xi-\xi_c)]+\delta/2\}}, \quad (24)$$

where $\Delta^2=1-\delta^2/4$. The phase difference $\Phi$ evolves according to expression

$$\Phi(\xi)=\text{sgn}\,\vartheta \arctan\frac{\Delta e^{2(\xi-\xi_c)}}{1+(\delta/2)e^{2(\xi-\xi_c)}}\pm\frac{\pi}{2}. \quad (25)$$

The expression (23)-(25) describe the exponentially decaying wave fronts. However some times the solitary waves can be decreasing as $\xi^{-2}$. The solutions found here are correct if $-2 < \delta < 2$. However, on the boundaries of this interval we have to refine behavior of the solitary waves. When $\delta \to -2$ the solution of the equations (21)-(22) take the form of the algebraic soliton [32]

$$|A|^2 = \frac{8}{|\vartheta|(1+\beta_1)[1+4(\xi-\xi_c)^2]}, \quad |B|^2 = \frac{8}{|\vartheta|(1-\beta_1)[1+4(\xi-\xi_c)^2]}. \quad (26)$$

Here $\beta_1$ corresponds with $\delta = -2$. On the other hand, the amplitudes of the solitary waves are equal to zero if $\delta \to +2$.

### 3.3. Robustness of the solitary waves in ANOWZA

To investigate of the solitary wave's stability the collision between two solitary waves has been simulated. To produce a collision between two solitary waves we used solutions of the equations (21) and (22) with parameters $\beta = 0.4$ and $\beta = -0.4$ as initial conditions for these equations. The pulse with $\beta = 0.4$ was located at $\zeta_c = 50$ and pulse with $\beta = -0.4$ was located at $\zeta_c = 0$. For simplicity the coupling constant $\gamma_1$ was set as unite, whereas $\gamma_2$ and $\gamma_3$ are assumed equal. Parameter $\gamma_2$ was varied from 0.001 to 0.2.

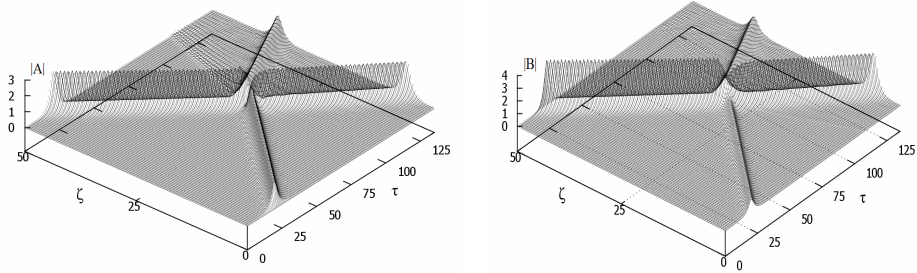

**Fig.5** Collision of two steady state pulses in the case of $\gamma_2 = 0.01$. Left panels are corresponding for PRI waveguides, right panels are corresponding for NRI waveguides [32].

It was found that the collision between counter propagating pulses is elastic for the coupling constant $\gamma_2$ that is taken from interval [0.001, 0.0075] (Fig.5). Little radiation appears where the coupling constant $\gamma_2$ is more then 0.0075. The amplitude of the radiation is increasing and at $\gamma_2 > 0.02$ the reflected wave appears in NRI waveguide as a result of the reflection of the incident solitary wave ($\beta = -0.4$) from solitary wave ($\beta = 0.4$) propagated in the opposite direction in ANOWZA (Fig.6).

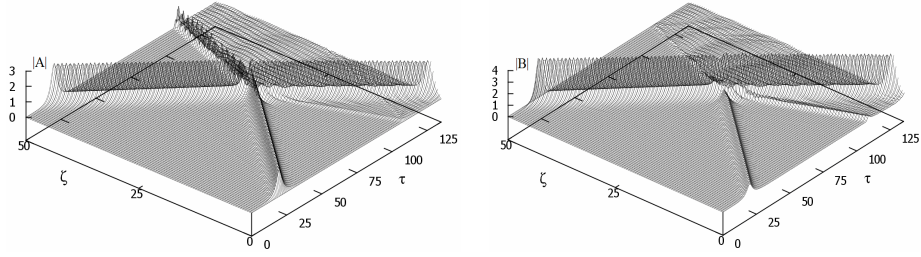

**Fig.6** Collision of two steady state pulses in the case of $\gamma_2 =0.06$. Left panels are corresponding for PRI waveguides, right panels are corresponding for NRI waveguides [32].

However, if the coupling constants belong to interval [0.08, 0.135] the steady state solitary waves are akin to elastic interacting waves. There is no radiation after collision, but the velocities of the scattered pulses can be strongly varied with respect of initial values.

The interaction of the incident solitary wave corresponding to $\beta = 0.4$ with quasi-harmonic wave has been considered. Initial incident pulse was located at $\zeta_c = 50$, the quasi-harmonic wave was generated at $\zeta = 0$ and it is characterized by frequency $\omega_{bg}$.

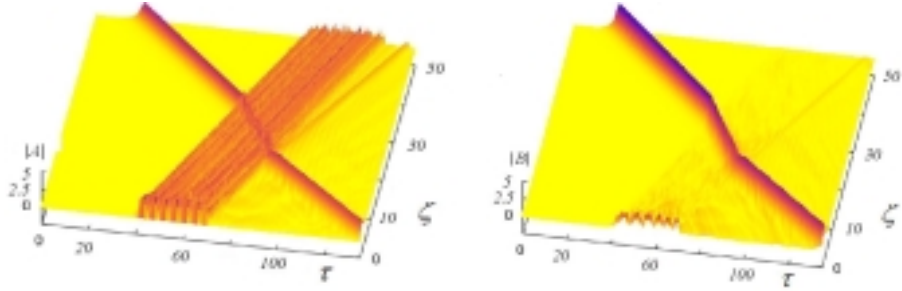

**Fig.7** Steady state pulses crossing the quasi-harmonic plateau like wave, $\omega_{bg} = 0.7$, $\gamma_2 =0.01$. Left panels are corresponding for PRI waveguides, right panels are corresponding for NRI waveguides.

It was found that initial steady state solitary wave is extremely prone to damage if the coupling constants $\gamma_2 = \gamma_3$ are less than 0.05. Fig. 7 shows the propagation of incident pulse trough quasi-harmonic wave having amplitude $f$ =1.5 and $\omega_{bg}$ = 0.7. The consideration of the case of more high frequency quasi-harmonic wave does not show any fundamental difference from the low frequency cases. Thus, the numerical simulation demonstrates robustness of these solitary waves.

## 4. Conclusion

Here the propagation of electromagnetic solitary wave in nonlinear coupler where one of the waveguide is fabricated from a material with a negative refractive index. The zigzag positive-negative coupled waveguide array is generalization of this simple device. These waveguide systems are present the situation in where forward and backward electromag-netic waves interact with each other. Alternate positive and negative refractive indexes result in the gap in linear wave spectrum. That is dissimilarity from a convenient couplers or waveguide array.

In nonlinear coupler and zigzag waveguides array the solitary forward and backward waves can be combined into single solitary wave, which is referred as gap soliton. It should be noted that the term *gap soliton* is often referred to as nonlinear pulses propagating in periodic structures. The waveguide structure considered here, however, is homogeneous. Hence, the existence of a gap soliton, and the bistability of continuous waves in an oppositely directed coupler represent new effects due to the positive-negative refraction phenomenon.

Recently the new intriguing properties of the twisted alternating waveguide array were predicted in [33]. It was show that forbidden zone in spectrum of linear waves can be controlled by the twist of waveguide array.

**Acknowledgements** We would like to thank N.M. Litchinitser, I.R. Gabitov, A.S. Desyatnikov and J.G. Caputo for enlightening discussions. AIM thanks the Department of Mathematics of University of Arizona, Laboratoire de Mathematiques, INSA de Rouen and the Nonlinear Physics Center, the Australian National University for the support and hospitality. This work was partially supported by NSF (grant DMS-0509589), ARO-MURI award 50342-PH-MUR and State of Arizona (Proposition 301), and by the Russian Foundation for Basic Research (grants No. 09-02-00701-a, 12-02-00561).